# Electron Self-energy in Pseudo-Hermitian Quantum Electrodynamics with a Maximal Mass *M*

*V.P.Neznamov\**
*RFNC-VNIIEF, Sarov, Russia*

## Abstract

The electron self-energy (self-mass) is calculated on the basis of the model of quantum field theory with maximal mass *M*, developed by V.G.Kadyshevsky et al. within the pseudo-Hermitian quantum electrodynamics in the second order of the perturbation theory.

In theory, there is the natural cut – off of large transmitted momentum in intermediate states because of presence of the universal mass *M*. As a result, the electron self – mass is finite and depends on the transmitted maximum momentum

$$k_{\max} = AMf(\mathbf{p}), \quad \left(\frac{M}{m} \gg 1, \quad A \ll \frac{M}{m}, f(\mathbf{p}) \ll \frac{M}{m}\right).$$

Two interpretations of the obtained results are possible at defined *M* and *A*.

The first interpretation allows confirming quantitatively the old concept of elementary particle mass sources defined by interaction of particles with self – gauge fields.

The second interpretation results in the possibility not to renormalize the mass (at least in the second order of perturbation theory) owing to the zero mass operator $\Sigma(p)$.

---

\* e-mail: neznamov@vniief.ru

## INTRODUCTION

V.G.Kadyshevsky and his colleagues in their papers [1 – 8] developed the M.A. Markov's idea about existence of a maximal elementary particle mass $M$ [9]. In these papers the existence of mass $M$ has been understood as a new fundamental principle of Nature similar to the relativistic and quantum postulates, which are put in the grounds of quantum field theory.

The condition of finiteness mass spectrum of elementary particles is expressed by introduction of relation

$$m \leq M , \qquad (1)$$

where mass parameter $M$ is a new physical constant. In the paper [9] by M.A.Markov,

$$M \cong m_{planck} = 10^{19} GeV .$$

In the papers [1 – 8], a new concept of a local quantum theory has been developed on the ground of relation (1) with construction of corresponding Lagrangians for boson and fermion fields. Relation (1) is satisfied by introduction of anti de Sitter space just as for satisfaction of relativistic conditions in due time was necessary the transition from a 3D space to a 4D Minkovsky space.

An analogue of the relativistic relation between energy and momentum of particle with fulfillment of relation (1) in anti de Sitter momentum space may be written as

$$p_0^2 - \mathbf{p}^2 + p_5^2 = M^2. \qquad (2)$$

The relation $p_0^2 - \mathbf{p}^2 = m^2$ is satisfied for particles of mass $m$ on the mass surface and apparently satisfaction of equality

$$p_5^2 = M^2 - m^2 . \qquad (3)$$

The Dirac equation properties for a fermion of mass $m$ on the mass surface $p_5 = \pm\sqrt{M^2 - m^2}$ were considered in [10]. The paper shows that free Hamiltonian and Hamiltonian with interaction are pseudo-Hermitian. The consistent quantum theory with final results coincident with the results obtained using the ordinary Dirac equation can be constructed when introducing the appropriate rules of use of pseudo-Hermitian Hamiltonians [11, 12, 13].

The eight-component spinors with four upper and four lower components which are the corresponding solutions of the four-component Dirac equation with different signs before the non-Hermitian summands with mass $m_1$, are introduced in the case of interaction between the Dirac particle and boson fields to provide the pseudo-Hermiticity requirement in [10].

The present paper goes into the calculation of the self-energy which is turned out finite and depends on the upper limit of integration $k_{\max} = AMf(\mathbf{p}), \left( \dfrac{M}{m} \gg 1, \ A \ll \dfrac{M}{m}, f(\mathbf{p}) \ll \dfrac{M}{m} \right)$ within the quantum electrodynamics on the basis of model with the maximal mass $M$ in the lowest order of perturbation theory.

Two interpretations of the obtained results are possible at defined $M$ and $A$.

The first interpretation allows confirming quantitatively the old concept of elementary particle mass sources defined by interaction of particles with self – gauge fields.

The second interpretation results in the possibility not to renormalize the mass (at least in the second order of perturbation theory) owing to the zero mass operator $\Sigma(p)$.

Given in the Section 1 is the Dirac equation Hamiltonian in anti de Sitter 5D space, basis functions of the Dirac particle free motion with orthonormality and completeness condition. The photon and electron-positron propagators are defined in the Sections 2 and 3. The self-energy of electron is calculated in the Sections 4. The discussion of results is given in the final Section.

1. **THE DIRAC EQUATION HAMILTONIAN WITH ELECTROMAGNETIC INTERACTION IN ANTI DE SITTER SPACE. FREE MOTION BASIS FUNCTIONS**

According to [8, 10], four Dirac equations with electromagnetic interaction and maximal mass $M$, can be written in momentum anti de Sitter space.

$$\begin{cases} p_0\psi_1(p,p_5) = (\boldsymbol{\alpha}\mathbf{p} + \beta\gamma_5(M-|p_5|) + \beta 2M\sin\frac{\mu}{2} + \alpha_\nu\mathcal{A}^\nu(p,p_5))\psi_1(p,p_5) \\ p_0\psi_2(p,p_5) = (\boldsymbol{\alpha}\mathbf{p} + \beta\gamma_5(M-|p_5|) - \beta 2M\sin\frac{\mu}{2} + \alpha_\nu\mathcal{A}^\nu(p,p_5))\psi_2(p,p_5); \end{cases}$$

$$\begin{cases} p_0\psi_3(p,p_5) = (\boldsymbol{\alpha}\mathbf{p} - \beta\gamma_5(M-|p_5|) + \beta 2M\sin\frac{\mu}{2} + \alpha_\nu\mathcal{A}^\nu(p,p_5))\psi_3(p,p_5) \\ p_0\psi_4(p,p_5) = (\boldsymbol{\alpha}\mathbf{p} - \beta\gamma_5(M-|p_5|) - \beta 2M\sin\frac{\mu}{2} + \alpha_\nu\mathcal{A}^\nu(p,p_5))\psi_4(p,p_5). \end{cases}$$ (4)

In (4) and below $\hbar = c = 1$, $\gamma^0 = \beta$, $\gamma^i = \beta\alpha^i$, $\gamma^5 = i\gamma^0\gamma^1\gamma^2\gamma^3$ are four-dimensional Dirac matrices; $\mu$ is defined by $\cos\mu = \sqrt{1-\frac{m^2}{M^2}}$, where $m$ – is the Dirac particle mass,

$\alpha^\nu = \begin{cases} 1, & \nu = 0 \\ \alpha^i, & \nu = i = 1,2,3 \end{cases}$; $\mathcal{A}^\nu(p,p_5)$ – integral expressions representing electromagnetic potentials $A^\nu(x_0, \mathbf{x})$ in anti de Sitter space. In equations (4) gauge is taken $\mathcal{A}_5 = 0$ [6, 9]; $p_5 = -|p_5|$ in the first two equations (4), $p_5 = |p_5|$ in equations for $\psi_3$ and $\psi_4$.

Equations (4) differ from each other only in signs before the terms with matrices $\beta\gamma_5$ and $\beta$. As for the physical consequences, equations (4) are equivalent to each other similarly to the ordinary Dirac equations with different signs before the mass term $m$.

$|p_5| = \sqrt{M^2 - m^2}$ and $M - |p_5| = m_1 = 2M\sin^2\frac{\mu}{2}$ in case the particle is on the mass surface $(p_0^2 - \mathbf{p}^2 = m^2)$. If we set $m_2 = 2M\sin\frac{\mu}{2}$, we have $m_2^2 - m_1^2 = m^2$.

If $m \ll M$, mass $m_2 \approx m(1+\frac{1}{8}\frac{m_2}{M^2})$ is close to a particle mass, and $m_1 \approx m\frac{m}{2M}$ is small.

It can be seen from the equation (4) that Hamiltonians are non-Hermitian because of the terms with matrices $\beta\gamma_5$. It can be shown that combination of two equations with different signs before the non-Hermitian terms leads to pseudo-Hermiticity of the combined Hamiltonian [10].

Pseudo-Hermiticity of the Hamiltonian allows consideration of the Hamiltonian in the context of non-Hermitian quantum mechanics formalism [11, 12, 13].

Consider two equations for $\psi_1(p,p_5)$ and $\psi_3(p,p_5)$ from the equation (4).

$$\begin{cases} p_0\psi_1(p,p_5) = (\boldsymbol{\alpha}\mathbf{p} + \beta\gamma_5(M-|p_5|) + \beta m_2 + \alpha_\nu\mathcal{A}^\nu(p,p_5))\psi_1(p,p_5); \; p_5 = -|p_5| \\ p_0\psi_3(p,p_5) = (\boldsymbol{\alpha}\mathbf{p} - \beta\gamma_5(M-|p_5|) + \beta m_2 + \alpha_\nu\mathcal{A}^\nu(p,p_5))\psi_3(p,p_5); \; p_5 = |p_5|. \end{cases}$$ (5)

Let us introduce the eight-component spinor $\phi(p,p_5)$ with four upper components being the solution for the equation (5) for $\psi_1(p,p_5)$ and with four lower components being the solution for the equation (5) for $\psi_3(p,p_5)$.

$$\phi(p,p_5) = \begin{pmatrix} \psi_1(p,p_5) \\ \psi_3(p,p_5) \end{pmatrix}.$$ (6)

We introduce also isotopic matrices $\tau_3 = \begin{pmatrix} I & 0 \\ 0 & -I \end{pmatrix}$ and $\tau_1 = \begin{pmatrix} 0 & I \\ I & 0 \end{pmatrix}$, which are effective in the isotopic space of the four upper and four lower components of spinor $\phi(p, p_5)$.

The equation (5) can be written as

$$p_0 \phi(p, p_5) = (\boldsymbol{\alpha}\mathbf{p} + \tau_3 \beta \gamma^5 (M - |p_5|) + \beta m_2 + \alpha_v \mathcal{A}^v(p, p_5)) \phi(p, p_5). \tag{7}$$

The equation (7) has two branches of possible values of $p_5 = \pm|p_5|$.

The pseudo-Hermiticity requirement $\rho H \rho^{-1} = H^\dagger$ can be written if we use $\rho = \tau_1 = \tau_1^{-1}$. Indeed

$$\tau_1 H_\phi \tau_1 = \tau_1 (\boldsymbol{\alpha}\mathbf{p} + \tau_3 \beta \gamma_5 (M - |p_5|) + \beta m_2 + \alpha_v \mathcal{A}^v(p, p_5)) \tau_1 = H_\phi^\dagger. \tag{8}$$

The equation (7) can be written in 4D Minkowsky space $(x) = (x^0, \mathbf{x})$ [10] in case of free motion $\left( \mathcal{A}_v(p, p_5) = 0, \ |p_5| = \sqrt{M^2 - m^2} \right)$

$$i \frac{\partial \phi(x)}{\partial x_0} = H_0 \phi(x) = (\boldsymbol{\alpha}\mathbf{p} + \tau_3 \beta \gamma_5 m_1 + \beta m_2) \phi(x). \tag{9}$$

Free motion basis functions for (9) correspond to the two positive-energy and to the two negative-energy solutions:

$$\phi_1^{(+)}(x, s) = U_s(T_3 = +1/2) e^{-ip_v x^v} = \begin{pmatrix} U_s(m_1) \\ 0 \end{pmatrix} e^{-ip_v x^v};$$

$$\phi_2^{(+)}(x, s) = U_s(T_3 = -1/2) e^{-ip_v x^v} = \begin{pmatrix} 0 \\ U_s(-m_1) \end{pmatrix} e^{-ip_v x^v};$$

$$\phi_1^{(-)}(x, s) = V_s(T_3 = +1/2) e^{ip_v x^v} = \begin{pmatrix} V_s(m_1) \\ 0 \end{pmatrix} e^{ip_v x^v}; \tag{10}$$

$$\phi_2^{(-)}(x, s) = V_s(T_3 = -1/2) e^{ip_v x^v} = \begin{pmatrix} 0 \\ V_s(-m_1) \end{pmatrix} e^{ip_v x^v};$$

$$U_s(\pm m_1) = \sqrt{\frac{E + m_2}{2E}} \begin{pmatrix} \varphi_s \\ \dfrac{\boldsymbol{\sigma}\mathbf{p} \mp m_1}{E + m_2} \varphi_s \end{pmatrix};$$

$$V_s(\pm m_1) = \sqrt{\frac{E + m_2}{2E}} \begin{pmatrix} \dfrac{\boldsymbol{\sigma}\mathbf{p} \mp m_1}{E + m_2} \chi_s \\ \chi_s \end{pmatrix}. \tag{11}$$

In (10) and (11), $\mathbf{p}$ and $E$ are the momentum and energy operators for a particle of mass $m$; $T_3 = \pm 1/2$ are the isotopic spin values; $U_s(T_3 = \pm 1/2)$, $V_s(T_3 = \pm 1/2)$ are the eight-component spinors; $U_s(\pm m_1)$, $V_s(\pm m_1)$ are the four-component spinors; $\varphi_s$ and $\chi_s$ are the two-component normalized Pauli spin functions.

The following relations of orthonormality and completeness are valid for $U_s(T_3 = \pm 1/2)$ and $V_s(T_3 = \pm 1/2)$, with regard to the Hamiltonian pseudo-Hermiticity of the equation (9):

$$U_s^\dagger(T_3 = \mp 1/2) \tau_1 U_{s'}(T_3 = \pm 1/2) = V_s^\dagger(T_3 = \mp 1/2) \tau_1 V_{s'}(T_3 = \pm 1/2) = \delta_{ss'};$$

$$U_s^\dagger(T_3 = \mp 1/2) \tau_1 V_{s'}(T_3 = \pm 1/2) = V_s^\dagger(T_3 = \mp 1/2) \tau_1 U_{s'}(T_3 = \pm 1/2) = 0; \tag{12}$$

$$\sum_s \left(U_s(T_3 = \pm 1/2)\right)_\alpha \left(U_s^\dagger(T_3 = \mp 1/2)\right)_\beta (\tau_1)_{\beta\gamma} = \left[1/2\left(1+\frac{H_0}{E}\right)1/2(1\pm\tau_3)\right]_{\alpha\gamma};$$

$$\sum_s \left(V_s(T_3 = \pm 1/2)\right)_\alpha \left(V_s^\dagger(T_3 = \mp 1/2)\right)_\beta (\tau_1)_{\beta\gamma} = \left[1/2\left(1-\frac{H_0}{E}\right)1/2(1\pm\tau_3)\right]_{\alpha\gamma}.$$

In expressions (10) – (12) $\alpha$, $\beta$, $\gamma$ are used for spinor indexes and $s$, $s'$ are used for spin indexes. In the further summation over spinor indexes the summation symbol and indexes may not be shown.

## 2. PHOTON PROPAGATOR

The Maxwell equation in momentum space for the free electromagnetic field can be written as [14]

$$-g^{\nu\lambda}k^2 A_\lambda(k) + \left(1 - \frac{1}{\xi}\right) k^\nu k^\lambda A_\lambda(k) = 0. \quad (13)$$

In (13), $g^{\nu\lambda}$ is the metric tensor ($g^{\nu\lambda} = diag[1,-1,-1,-1]$), $\xi$ is an arbitrary constant ($\xi = 1$ corresponds to the Feynman gauge, the case $\xi = 0$ is called the Landau gauge).

According to [5], [8], transformation to anti de Sitter space as well as appearance of mass $M$, variable $k_5$, potential $A_5(k, k_5)$ result in the transformation of (13) into the following set of equations

$$(k_5 - M) A_\lambda(k, k_5) + \left(1 - \frac{1}{\xi}\right) k_\lambda A_5(k, k_5) = 0;$$

$$(k_5 + M) A_5(k, k_5) + k^\lambda A_\lambda(k, k_5) = 0. \quad (14)$$

Equations (14) can be written as

$$\left(g^{\nu\lambda}(k_5 - M) + \left(1 - \frac{1}{\xi}\right) \frac{k^\nu k^\lambda}{k^5 + M}\right) A_\lambda(k, k_5) = 0. \quad (15)$$

It follows from (15) that the inverse operator or propagator in momentum space can be written as

$$D_{\mu\nu}(k, k_5) = -\frac{g_{\mu\nu}(k_5 + M)}{k^2} - (\xi - 1)\frac{(k_5 + M)k_\mu k_\nu}{k^4}. \quad (16)$$

Further we use the Feynman gauge $\xi = 1$.

In [8], mostly for technical reasons, when developing the theory the authors convert to its Euclidean formulation due to the substitution of $k_0 \to ik_4$. In this case, there is a transition to de Sitter space, instead of anti de Sitter space (2)

$$-k_1^2 - k_2^2 - k_3^2 - k_4^2 + k_5^2 = -\mathbf{k}^2 + k_5^2 = M^2. \quad (17)$$

It is obvious that:

$$k_5 = \pm\sqrt{M^2 + \mathbf{k}^2}. \quad (18)$$

Further we use the Euclidean formulation of theory (17), (18) for unequivocal record of propagator (16) in configuration space $(\mathbf{x}, x_5)$. In this formulation

$$D_{\mu\nu}^{Eucl}(\mathbf{k}, k_5) = \frac{\delta_{\mu\nu}(k_5 + M)}{\mathbf{k}^2} + (\xi - 1)\frac{(k_5 + M)\mathbf{k}_\mu \mathbf{k}_\nu}{\mathbf{k}^4}. \quad (19)$$

The propagation function in the configuration space $(\mathbf{x}, x_5)$ is written as

$$D_{\mu\nu}(\mathbf{x}, x_5) = \frac{i\delta_{\mu\nu}}{2(2\pi)^4} \int \frac{d^4\mathbf{k}}{\mathbf{k}^2} e^{i\mathbf{k}\mathbf{x}} \frac{1}{|k_5|} \left[ \left( |k|_5 + M \right) e^{-i|k_5||x_5|} - \left( |k|_5 - M \right) e^{i|k_5||x_5|} \right]. \tag{20}$$

As the electromagnetic field action does not depend on $x_5$-coordinate [5], [8], it can be further set to zero $(x_5 = 0)$. We also take into account that $|k_5| = \sqrt{M^2 + \mathbf{k}^2}$. Then

$$D_{\mu\nu}(\mathbf{x}) = \frac{i}{(2\pi)^4} \int d^4\mathbf{k} D_{\mu\nu}^{Eucl}(\mathbf{k}) e^{i\mathbf{k}\mathbf{x}} = \frac{i\delta_{\mu\nu}}{(2\pi)^4} \int \frac{d^4\mathbf{k}}{\mathbf{k}^2} \frac{M}{\sqrt{M^2 + \mathbf{k}^2}} e^{i\mathbf{k}\mathbf{x}}. \tag{21}$$

Expression (21) differs from the standard photon propagator in the factor $\dfrac{M}{\sqrt{M^2 + \mathbf{k}^2}}$ in the integrand. This factor is the new factor effecting on the electron self-energy estimated in Section 4.

### 3. ELECTRON-POSITRON PROPAGATOR

The Dirac equation for the free motion is estimated in Section 1 (see (7)) and is written in momentum anti de Sitter space as

$$(p_0 - \boldsymbol{\alpha}\mathbf{p} - \tau_3\beta\gamma^5(M - |p_5|) - \beta m_2)\phi(p, p_5) = 0. \tag{22}$$

The equation (22) defines the kind of propagator in Euclidean formulation by analogy with previous Section

$$S^{Eucl}(\mathbf{p}, p_5) = \frac{1}{ip_4 - \boldsymbol{\alpha}\mathbf{p} - \tau_3\beta\gamma^5(M - |p_5|) - \beta m_2}. \tag{23}$$

The propagation function in the configuration space $(\mathbf{x}, x_5)$ is given by

$$S(\mathbf{x}, x_5) = \frac{iM}{2(2\pi)^4} \int \frac{d^4\mathbf{p}}{|p_5|} S^{Eucl}(\mathbf{p}, p_5) e^{i\mathbf{p}\mathbf{x}} \left[ e^{-i|p_5||x_5|} + e^{i|p_5||x_5|} \right]. \tag{24}$$

When $x_5 = 0$

$$S(\mathbf{x}) = \frac{iM}{(2\pi)^4} \int \frac{d^4\mathbf{p}}{|p_5|} S^{Eucl}(\mathbf{p}, p_5) e^{i\mathbf{p}\mathbf{x}} =$$
$$= -\frac{i}{(2\pi)^4} \int d^4\mathbf{p} \frac{1}{2}\left(1 + \frac{\sqrt{M^2 - m^2}}{\sqrt{M^2 + \mathbf{p}^2}}\right) \frac{ip_4 + \boldsymbol{\alpha}\mathbf{p} + \tau_3\beta\gamma_5\left(M - \sqrt{M^2 + \mathbf{p}^2}\right) + \beta m_2}{\mathbf{p}^2 + m^2} e^{i\mathbf{p}\mathbf{x}}. \tag{25}$$

Equation (23) and (25) define the kind of electron-positron propagator in momentum and configuration space.

In equations (17) – (25) and below $\mathbf{x}, \mathbf{k}, \mathbf{p}$ are 4D vectors with components $n = 1, 2, 3, 4$ and metric $diag[-1, -1, -1, -1]$.

### 4. ELECTRON SELF-ENERGY

In theory under consideration an action does not depend on coordinate $x_5$ [5], [8]. In addition, the kind of interaction Lagrangian in pseudo-Hermitian electrodynamics with mass $M$ coincides with standard one [15]

$$L_{int} = -ei \int d^4\mathbf{x} \psi^\dagger(\mathbf{x}) \alpha_\mu A^\mu(\mathbf{x}) \psi(\mathbf{x}). \tag{26}$$

Hence it follows that we can use the perturbation theory with the Feynman standard rules for calculation of concrete processes of pseudo-Hermitian quantum electrodynamics. But in this case it should be use the photon and the electron-positron propagators, which are given in expressions (19), (21), (23), (25).

Electron self-energy is estimated in the second order of perturbation theory. The appropriate diagram of process is given in Figure.

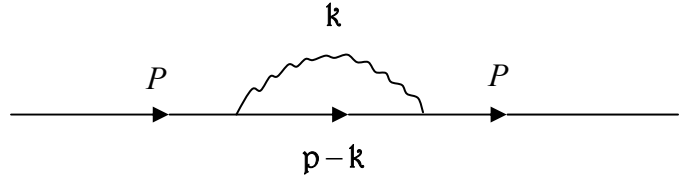

Electron self-energy.

The appropriate element of scattering matrix with self-energy operator (self-mass operator) $\Sigma(p)$ can be written as

$$S_{fi} = -U_{s_f}^{\dagger}(T_3 = \mp 1/2)\tau_1 \Sigma(p) U_{s_i}(T_3 = \pm 1/2). \tag{27}$$

In expression (27) the mass operator is between the corresponding eight-component spinors with matrix $\tau_1$ and with opposite signs of the isotopic spin third component $T_3$ in input and output Feynman lines (see discussion in Section 2).

Expression (27) implies that an electron is on the mass surface $(p_0^2 - \mathbf{p}^2 = m^2)$. In this case, expressions (24), (25) give $|p_5| = \sqrt{M^2 - m^2}$ and $(M - |p_5|) = 2M \sin^2 \frac{\mu}{2} = m_1$.

Expression for self-energy operator (mass operator) $\Sigma(p)$ corresponding to the diagram of the Figure is

$$\Sigma(p) = (-ie)^2 \int \frac{d^4 k}{(2\pi)^4} i D_{\mu\nu}^{Eucl}(k) \alpha_\mu i S^{Eucl}(p-k) \alpha^\nu =$$

$$= -\frac{e^2}{(2\pi)^4} \int \frac{d^4 k}{k^2} \frac{M}{\sqrt{M^2 + k^2}} \alpha_\nu \frac{1}{i(p_4 - k_4) - \boldsymbol{\alpha}(\mathbf{p}-\mathbf{k}) - \tau_3 \beta\gamma m_1 - \beta m_2} \alpha^\nu. \tag{28}$$

For computational convenience in integral (28), the inverse Wick rotation has been done as well as the transition from Euclidean space to Minkowski space with metric tensor

$$g^{\nu\lambda} = diag[1,-1,-1,-1] \quad (d^4 k = dk_0 d\mathbf{k}, \ k^2 = k_0^2 - \mathbf{k}^2, \ p_0 = ip_4).$$

Expression (28) becomes

$$\Sigma(p) = \frac{-ie^2}{(2\pi)^4} \int \frac{d^4 k}{k^2} \frac{M}{\sqrt{M^2 - k^2}} \alpha_\nu \frac{p_0 - k_0 + \boldsymbol{\alpha}(\mathbf{p}-\mathbf{k}) + \tau_3 \beta\gamma^5 m_1 + \beta m_2}{(p_0 - k_0)^2 - (\mathbf{p}-\mathbf{k})^2 - m^2} \alpha^\nu. \tag{29}$$

The two features distinguish the expression (29) for mass operator from the standard expression in quantum electrodynamics (see, for example, [16]). The first feature is the occurrence of two terms with masses $m_1$, $m_2$, one of which is non-Hermitian, in the term instead of the expression $\beta m$.

The results of the previous work [10] show that quantum theory final results for the particles on mass surface agree with standard results subject to the proper handling of pseudo-Hermitian Hamiltonian.

The second feature of (28) – is the factor $\dfrac{M}{\sqrt{M^2 - k^2}}$, imposing the restriction on 4D momentum integration domain $k$ ($k^2 = k_0^2 - \mathbf{k}^2 \leq M^2$). In our case this restriction results in the finite expression for electron self-energy.

The notation in (28) $\tilde{p}_0 = (m^2 + (\mathbf{p} - \mathbf{k})^2)^{1/2}$ and summation over $v$, give

$$-i\Sigma(p) = -\frac{2e^2}{(2\pi)^4} \int_{k_0^2 \leq M^2 + \mathbf{k}^2} \frac{dk_0 d\mathbf{k}}{(k_0 - |\mathbf{k}|)(k_0 + |\mathbf{k}|)} \frac{M}{\sqrt{M^2 + \mathbf{k}^2 - k_0^2}} \times$$
$$\times \frac{k_0 - p_0 + \boldsymbol{\alpha}(\mathbf{p} - \mathbf{k}) + 2\tau_3 \beta\gamma^5 m_1 + 2\beta m_2}{(k_0 - \tilde{p}_0 - p_0)(k_0 + \tilde{p}_0 - p_0)}. \tag{30}$$

Consider $M \cong m_{planck} = 10^{19} GeV$, then $k_0^2 \leq \mathbf{k}^2 + M^2$ allows $k_0$ contour integration in (30), if we use theorem of residues.

$$-i\Sigma(p) = \frac{ie^2}{2(2\pi)^3} \int d\mathbf{k} \left[ \frac{-p_0 + \kappa + \boldsymbol{\alpha}(\mathbf{p} - \mathbf{k}) + 2\tau_3 \beta\gamma^5 m_1 + 2\beta m_2}{\kappa(p_0 - \kappa - \tilde{p}_0)(p_0 - \kappa + \tilde{p}_0)} + \right.$$
$$+ \frac{-p_0 - \kappa + \boldsymbol{\alpha}(\mathbf{p} - \mathbf{k}) + 2\tau_3 \beta\gamma^5 m_1 + 2\beta m_2}{\kappa(p_0 + \kappa - \tilde{p}_0)(p_0 + \kappa + \tilde{p}_0)} +$$
$$+ \frac{M}{\sqrt{M^2 + \mathbf{k}^2 - (\tilde{p}_0 + p_0)^2}} \frac{\tilde{p}_0 + \boldsymbol{\alpha}(\mathbf{p} - \mathbf{k}) + 2\tau_3 \beta\gamma^5 m_1 + 2\beta m_2}{\tilde{p}_0(\tilde{p}_0 - \kappa + p_0)(\tilde{p}_0 + \kappa + p_0)} +$$
$$\left. + \frac{M}{\sqrt{M^2 + \mathbf{k}^2 - (\tilde{p}_0 - p_0)^2}} \frac{-\tilde{p}_0 + \boldsymbol{\alpha}(\mathbf{p} - \mathbf{k}) + 2\tau_3 \beta\gamma^5 m_1 + 2\beta m_2}{\tilde{p}_0(\tilde{p}_0 - \kappa - p_0)(\tilde{p}_0 + \kappa - p_0)} \right]. \tag{31}$$

Let us assume $\kappa = |\mathbf{k}|$ in (31).

Radicand of the two last terms in (31) are

$$M^2 + \mathbf{k}^2 - (\tilde{p}_0 \pm p_0)^2 = M^2 - 2(p_0^2 \pm \tilde{p}_0 p_0 - \mathbf{pk}) \geq 0. \tag{32}$$

In (32) $\tilde{p}_0 = (p_0^2 + \mathbf{k}^2 - 2\mathbf{pk})^{1/2}$.

The positivity condition (32) results in natural constraint of integration limit of $\mathbf{k}$ momentum. This condition for the upper signs in (32) for the electron at rest gives the upper limit of $|\mathbf{k}|$ integration

$$\kappa_{max} = \frac{M\sqrt{M^2 - 4m^2}}{2m} \approx \frac{M^2}{2m}. \tag{33}$$

The last equality accounts for the condition $m \ll M$.

If $M \approx 10^{19} GeV$, then $\kappa_{max}$ in (33) is a large but a finite quantity $\kappa_{max} = 10^{41} GeV$.

Let us assume $\kappa_{max} = AM$, $A \ll \dfrac{M}{m}$ to estimate (31) for mass operator $\Sigma(p)$ at electron energies $p_0 \ll M$.

In this case the radicand positivity (32) is provided in advance, and besides, square roots in (31) can be expanded by the smallness of the second term in (32), confining to the three expansion terms:

$$\frac{M}{\sqrt{M^2 + \mathbf{k}^2 - (\tilde{p}_0 \pm p_0)^2}} \approx \left(1 + \frac{p_0^2 \pm \tilde{p}_0 p_0 - \mathbf{pk}}{M^2} + \frac{3(p_0^2 \pm \tilde{p}_0 p_0 - \mathbf{pk})^2}{2M^4}\right). \tag{34}$$

Accounting for the next following expansion terms in (34) leads to extremely small contribution in $\Sigma(p)$, proportional to the relation degree of $\dfrac{m}{M}$.

With $\alpha = \dfrac{e^2}{4\pi}$ to denote the constant of thin structure, the expression (31) can be written in view of (34) as

$$\Sigma(p) = \frac{\alpha}{4\pi^2} \int d\mathbf{k} \left\{ \frac{1}{\kappa(\tilde{p}_0 + \kappa + p_0)} - \frac{1}{\kappa(\tilde{p}_0 + \kappa - p_0)} + \right.$$
$$+ (\boldsymbol{\alpha}(\mathbf{p}-\mathbf{k}) + 2\tau_3 \beta \gamma^5 m_1 + 2\beta m_2) \left[ \frac{1}{\tilde{p}_0 \kappa} \left( \frac{1}{\tilde{p}_0 + \kappa + p_0} + \frac{1}{\tilde{p}_0 + \kappa - p_0} \right) - \right. \tag{35}$$
$$\left. - \frac{1}{M^2 \tilde{p}_0} - \frac{3}{2M^4} \frac{(p_0^2 - \mathbf{p}\mathbf{k})}{\tilde{p}_0} \right] - (\boldsymbol{\alpha}\mathbf{p} + \tau_3 \beta \gamma^5 m_1 + \beta m_2) \frac{3\tilde{p}_0}{2M^4} \right\}.$$

The Dirac approach [17] can be used when integrating (35). The system with zero $p_1$ and $p_2$ is taken as a system of coordinates, i.e. $k_3$ direction coincides with $\mathbf{p}$ direction.

Next, a new variable is introduced $\omega = \tilde{p}_0 + |\mathbf{k}|$. Then $d\mathbf{k} = \dfrac{|\mathbf{k}|\tilde{p}_0}{\omega} d\omega dk_3 d\varphi$, where $\varphi$ - is an azimuth angle.

The terms of the integrand in (35) which does not depend on the parameter $M$, makes a standard contribution to the electron self-energy taking into account the limit of integration $\kappa_{max} = AM \gg m$ (see, for example, [17])

$$\sum\nolimits_1 (p) = (\tau_3 \beta \gamma^5 m_1 + \beta m_2) \frac{\alpha}{\pi} \left( \frac{3}{2} \ln\left(\frac{2AM}{m}\right) - \frac{1}{4} \right). \tag{36}$$

The other terms in (35) make the following contribution to $\Sigma(p)$:

$$\sum\nolimits_2 (p) = -(\tau_3 \beta \gamma^5 m_1 + \beta m_2) \frac{\alpha}{\pi} \left( A^2 + \frac{3}{8} A^4 + o\left(\frac{m}{M}\right) \right) - \boldsymbol{\alpha}\mathbf{p} \frac{\alpha}{\pi} \left( \frac{A^2}{6} + \frac{A^4}{2} + o\left(\frac{m}{M}\right) \right). \tag{37}$$

In (37), $o\left(\dfrac{m}{M}\right)$ are small values proportional to the degrees of $\dfrac{m}{M}$.

The mean of operators $\tau_3 \beta \gamma^5 m_1 + \beta m_2$ and $\boldsymbol{\alpha}\mathbf{p}$ (see (27)) is

$$\left\langle i; T_3 = \mp 1/2 \,|\, \tau_1 (\tau_3 \beta \gamma_5 m_1 + \beta m_2) \,|\, i; T_3 = \pm 1/2 \right\rangle = \frac{m^2}{p_0}; \tag{38}$$

$$\left\langle i; T_3 = \mp 1/2 \,|\, \tau_1 \boldsymbol{\alpha}\mathbf{p} \,|\, i; T_3 = \pm 1/2 \right\rangle = \frac{\mathbf{p}^2}{p_0}. \tag{39}$$

It follows from (37) - (39) that the second term in (37) proportional to $\boldsymbol{\alpha}\mathbf{p}$, is relativistically noninvariant. The analysis of (38), (39) shows that this term can be eliminated with fitted function being specially defined $f(\mathbf{p}) \ll \dfrac{M}{m}$, and the upper limit of integration $AM$ being substituted for $AMf(\mathbf{p})$. For example, function $f(\mathbf{p})$ can be fitted when setting the sum to zero with the corresponding expansion terms in powers of $\dfrac{\mathbf{p}^2}{m^2}$ when calculating $\Sigma(p)$ accounting for (38) and (39).

Then, in view of $\frac{\mathbf{p}^4}{m^4}$ expansion terms when $A=1$, the function $f(\mathbf{p})$ can be represented as

$$f(\mathbf{p}) = \left(1 - \frac{\mathbf{p}^2}{3m^2} + \frac{\mathbf{p}^4}{6m^4} + \ldots\right). \tag{40}$$

For arbitrary value $A$, the coefficients $a$ and $b$ in expansion $f(\mathbf{p}) = \left(1 + a\frac{\mathbf{p}^2}{m^2} + b\frac{\mathbf{p}^4}{m^4} + \ldots\right)$, are

$$a = \frac{\frac{A^2}{6} + \frac{A^4}{2}}{\frac{3}{2} - 2A^2 - \frac{3}{2}A^4}; \quad b = \frac{\frac{3}{4}a^2 + a^2 A^2 + \frac{9}{4}a^2 A^4 + \frac{1}{3}aA^2 + 2aA^4}{\frac{3}{2} - 2A^2 - \frac{3}{2}A^4}. \tag{41}$$

The procedure given is the relaivistically invariant as the function $f(\mathbf{p})$ is properly changed at change of the reference system. For the electron at rest $(\mathbf{p} = 0)$ $f(\mathbf{p}) = 1$.

All the above considered, the resultant expression for $\sum(p)$ is

$$\sum(p) = \sum\nolimits_1(p) + \sum\nolimits_2(p) = \frac{\alpha}{\pi}\left(\frac{3}{2}\ln\frac{2AM}{m} - \frac{1}{4} - A^2 - \frac{3}{8}A^4 + o\left(\frac{m}{M}\right)\right)(\tau_3 \beta \gamma^5 m_1 + \beta m_2). \tag{42}$$

Mean value of operator (see (27)) defines the expression for the final self-energy of electron with the upper limit of integration $k_{max} = AMf(\mathbf{p})$, $\left(\frac{M}{m} \gg 1, \ A \ll \frac{M}{m}, f(\mathbf{p}) \ll \frac{M}{m}\right)$.

$$\Delta E^{(2)} = \frac{m}{E}\Delta m^{(2)}; \quad \Delta m^{(2)} = m\frac{\alpha}{\pi}\left(\frac{3}{2}\ln\frac{2AM}{m} - \frac{1}{4} - A^2 - \frac{3}{8}A^4\right). \tag{43}$$

The sign $^{(2)}$ points out to the fact that the electron self-energy is defined in the second order of perturbation theory (see Figure).

**DISCUSSION OF RESULTS**

Shown here is the finiteness of the expression for the electron self-energy (42), (43) in the second order perturbation theory in the in the framework of pseudo-Hermitian quantum electrodynamics with maximal mass $M$. The universal mass $M$ leads to fundamental length $\frac{\hbar}{Mc}$ and sets limit of transmitted momentums in intermediate states.

For $M = 10^{19} GeV$ and $A = 1$, the self-mass is $\Delta m^{(2)} \approx 0,17m$ (see (42), (43)), it forms a notable part of the electron mass. The self – mass is comparable to the electron mass for a large but finite mass $M = 2 \cdot 10^{121} GeV$. Most likely this value of $M$ will reduce if the contribution of gravitation forces to the self-mass of electron is taken into account because the gravitational interaction is compared on the force to other interactions already for energies of $\sim 10^{18} GeV$ [18]. For example, for 50% contribution of gravitational interaction to the self-mass on condition that it is comparable with the electron mass the necessary value of $M$ is about $10^{59} GeV$.

The estimations assume the negligible contribution to the electron self-mass of weak interactions and high order effects of perturbation theory.

Within this approach, electron mass as well as mass of other elementary particles will be completely defined by interaction effects with self-fields. The approach is not new, it has been discussed by different researchers (see, for example, [19] and references). Besides, as stated in [19], the given approach stipulates automatically the presence of the neutrino mass of different generations by weak interactions with gauge self-fields. On the other hand, an expected low neutrino mass $(1-10eV)$ shows that there is a small contribution of weak interactions to lepton self-mass.

The expression (42) shows the other possibility for the electron self-energy. If $M = 10^{19} GeV$ and $A \approx 3,62$, operator $\Sigma(p)$ is zero. In this case there is no necessity to renormalize the electron mass. The electron mass is introduced into the theory from the very beginning from the outside and maintains the value up to the end of any given computations. The similar coefficients $A$ for the leptons of other generations of the Standard model for $M = 10^{19} GeV$, where $\Sigma(p) = 0$, give $A$=3,54 (muons - $m_\mu = 106 MeV$) and $A$=3,48($\tau$ - leptons - $m_\tau = 1800 MeV$). To answer the question of possibility of the identical values of $M$ and $A$, which give zero value of $\Sigma(p)$ for the leptons of three generations of the Standard model, it is necessary to make the corresponding calculations of the self-energy in the next (fourth) order of perturbation theory.

We note finally that owing to a large mass $M \geq 10^{19} GeV$, the radiation corrections earlier computed in the Standard model and particularly in quantum electrodynamics, will agree with similar calculated data in the theory with maximal mass M with an accuracy much more than present-day experimental data accuracy.

**Acknowledgements.** The author would like to thank V.G.Kadyshevsky for the useful discussions, advices and criticism.